\newcommand{\arxiv}{1}

\ifdefined\arxiv
	\documentclass[]{article} \else
	\documentclass[]{aircc} \fi

\usepackage{cite}

\usepackage{cmap}
\usepackage[T1]{fontenc}

\usepackage{graphicx}
\usepackage{silence}
\usepackage{subcaption}
\usepackage{xcolor}

\usepackage{amsmath} \usepackage{amssymb} 

\usepackage{bm} \usepackage{url}

\usepackage{calc}

\usepackage{amsthm}

\usepackage{xspace}

\definecolor{CLR1_}{HTML}{1b9e77}
\definecolor{CLR2_}{HTML}{d95f02}
\definecolor{CLR3_}{HTML}{7570b3}
\colorlet{CLR1}{CLR1_!50!white}
\colorlet{CLR2}{CLR2_!100!black}
\colorlet{CLR3}{CLR3_!50!black}

\colorlet{ALT1}{lightgray!50!white}
\colorlet{ALT2}{white!50!white} 
\usepackage{hyperref}

\newcommand{\vx}[1]{\overline{\bm{#1}}}
\newcommand{\AVG}{\bar} 

 \DeclareMathOperator*{\argmax}{arg\,max}

\newcommand{\UB}{\hat}

\newcommand{\UserIndex}{n}

\newcommand{\N}{^\UserIndex} \newcommand{\NI}{^\UserIndex_\FlowIndex}

\newcommand{\weight}{\omega} \newcommand{\Weight}{\weight{\NI}}  \newcommand{\VxWeightT}{\vx{\weight}(t)} \newcommand{\weightB}{\phi}
\newcommand{\WeightB}{\weightB{\NI}}

\newcommand{\capac}{C} \newcommand{\Capac}{C\NI} \newcommand{\VxCap}{\vx{C}}  \newcommand{\MaxCapac}{\hat{C}{\N}}

\newcommand{\rhoG}{\rho_{g}} \newcommand{\rhoM}{\rho_{M}} \newcommand{\RhoG}{\rhoG\NI}
\newcommand{\RhoM}{\rhoM\NI}

\newcommand{\TBTokens}{k}
\newcommand{\tbgtokens}{\TBTokens_g}
\newcommand{\tbmtokens}{\TBTokens_M}
\newcommand{\TBGTokens}{\tbgtokens\NI}
\newcommand{\TBMTokens}{\tbmtokens\NI}

\newcommand{\TBSigma}{\sigma}
\newcommand{\tbgsigma}{\TBSigma_g}
\newcommand{\tbmsigma}{\TBSigma_M}
\newcommand{\TBGSigma}{\tbgsigma\NI}
\newcommand{\TBMSigma}{\tbmsigma\NI}

\newcommand{\rate}{\capac} 			\newcommand{\Rate}{\Capac}
\newcommand{\sysState}{\mathcal{S}}	
\newcommand{\queue}{Q}				\newcommand{\Queue}{\queue{\NI}}

\newcommand{\hol}{\Gamma}			\newcommand{\HOL}{\hol{\NI}}

\newcommand{\util}{u}				\newcommand{\Util}{\util{\NI}}

\newcommand{\qosT}{\UB{T}} 			\newcommand{\QOST}{\qosT{\NI}}
\newcommand{\qosPVio}{\varepsilon}

		\newcommand{\AvgRate}{\AVG{\rate}\NI}

\newcommand{\BigOh}[1]{\mathcal{O}(#1)}
\newcommand{\RateRegion}{\mathcal{R}}
\newcommand{\NatNum}{\mathbb{N}}

\newcommand{\Mbit}[1]{#1Mbit} \newcommand{\Mbps}[1]{#1Mbps}
\newcommand{\secnd}[1]{#1\text{s}} 

\newcommand{\mycomment}[1]{}
\newcommand{\rangeII}[2]{\left[#1,#2\right]}
\newcommand{\rangeIE}[2]{\left[#1,#2\right[}
\newcommand{\rangeEI}[2]{\left]#1,#2\right]}

\usepackage{pgfplots} \usepackage{pgfplotstable} \pgfplotstableset{format=file,header=true,col sep=tab} 

\usetikzlibrary{plotmarks}

\pgfplotscreateplotcyclelist{my color list}{black,solid,mark=square\\blue,solid,mark=otimes\\red,solid,mark=triangle\\green,solid,mark=diamond\\purple,solid,mark=star\\\\
	blue,densely dashed,mark=*\\red,densely dashed,mark=square*\\green,densely dashed,mark=otimes*\\purple,densely dashed,mark=triangle*\\orange,densely dashed,mark=diamond*\\black,densely dashed,mark=star*\\}

\usetikzlibrary{pgfplots.colorbrewer}
\usepgfplotslibrary{colorbrewer}
\definecolor{dark2-7}{HTML}{a6761d}
\pgfplotscreateplotcyclelist{schedulers}{mark options=solid,line width=1pt, mark size=2pt,densely dotted,mark=o,\\mark options=solid,line width=1pt, mark size=2pt,densely dotted,mark=otimes\\mark options=solid,line width=1pt, mark size=2pt,densely dotted,mark=triangle\\mark options=solid,line width=1pt, mark size=2pt,densely dotted,mark=diamond\\mark options=solid,line width=1pt, mark size=2pt,densely dotted,mark=star\\mark options=solid,line width=1pt, mark size=2pt,solid,mark=square,dark2-7\\}

\usetikzlibrary{calc}

\usepgfplotslibrary{external} 

\ifdefined\arxiv
	\pgfplotsset{compat=1.14} \else
	\pgfplotsset{compat=1.16}
\fi

\usepackage[nolist]{acronym}
\newacro{QoS}{Quality of Service}
\newacro{QoE}{Quality of Experience}
\newacro{NUM}{Network Utility Maximization}
\newacro{MDV}{Minimal Delay Violation}
\newacro{MD}{Min-Delay}
\newacro{MW}{Max-Weight}
\newacro{DMW}{Delay-based Max-Weight}
\newacro{EDF}{Earliest Deadline First}
\newacro{PLR}{Packet Loss Ratio}
\newacro{CBR}{Constant Bit-Rate}

\newacro{HM}{Harmonic Mean}
\newacro{AM}{Arithmetic Mean}
\newacro{AG}{Geometric Mean}

\newacro{M-LWDF}{Modified Largest Weighted Delay First}
\newacro{EXP/PF}{Exponential/PF}
\newacro{MDU}{Maximal Delay Utility}
\newacro{DSL}{Digital Subscriber Line}
\newacro{TBRM}{Token Bucket Rate Modifier}
\newacro{GEO}{Geosynchronous earth orbit}
\newacro{MEO}{Medium Earth Orbit}
\newacro{VoIP}{Voice over IP}
\newacro{AC}{Admission Control}
\newacro{OSI}{Open Systems Interconnection}
\newacro{ONT}{Optical Network Terminal}
\newacro{4GBB}{4GBB}
\newacro{5GBB}{5GBB}
\newacro{OFDMA}{}

\newacro{IPDV}{Instantaneous Packet Delay Variance}
\newacro{PHB}{Per-Hop Behaviour}
\newacro{DSCP}{DiffServ Code Point}
\newacro{DiffServ}{Differentiated Services}
\newacro{IntServ}{Integrated Services}
\newacro{MPLS}{Multiprotocol Label Switching}
\newacro{E2E}{End-to-end delay}
\newacro{HOL}{Head-of-line}
\newacro{MSBs}{Most Significant Bits}
\newacro{WDRR}{Weighted Deficit Round-Robin}
\newacro{DRR}{Deficit Round-Robin}
\newacro{RR}{Round-Robin}

\newacro{iport}{ingress port}
\newacro{eport}{egress port}

\newacro{MBAC}{Measurement Based Admission Control}
\newacro{MSS}{Maximum Segment Size}
\newacro{TCP}{Transmission Control Protocol}
\newacro{UDP}{User Datagram Protocol}

\newacro{ANN}{Artificial Neural Network}
\newacro{PON}{Passive Optical Network}
\newacro{XG-PON}{10G-Passive Optical Network}
\newacro{ONU}{Optical Network Unit}
\newacro{CO}{Central Office}
\newacro{OLT}{Optical Line Terminal}
\newacro{DBA}{Dynamic Bandwidth Allocation}
\newacro{DF}{Demand Forecasting}
\newacro{RR}{Round Robin}
\newacro{TCONT}{Traffic Container}
\newacro{FTTH}{Fiber to the Home}
\newacro{ST}{Satellite Terminal}
\newacro{DBRu}{Dynamic Bandwidth Report upstream}
\newacro{DF-DBA}{Demand Forecasting Dynamic Bandwidth Allocation}

\newacro{LMS}{Least Mean Square}
\newacro{NLMS}{Normalized Least Mean Square}
\newacro{LMMSE}{Linear Minimum Mean Square Error}
\newacro{MA}{Moving Average}
\newacro{AR}{Autoregressive}
\newacro{ARIMA}{Autoegressive Integrated Moving Average}

\newacro{2g2u}{2 groups, 2 users per group}
\newacro{3g2u}{3 groups, 2 users per group}
\newacro{2g3u}{2 groups, 3 users per group} 

\newcommand{\MIIYLABEL}{Avg exc. bits/win (Mbp/win)}
\newcommand{\MIIIYLABEL}{Avg Busy Period (window)}
\newcommand{\MVYLABEL}{\% non-conforming slots}

\pgfplotsset{cycle list name=my color list}

\newcommand{\ADDHIST}[3]	{
	\addplot+
		[mark options={scale=1},mark indices={1,10,20,30,40,50,60,70,80,90,100}] 
		table[x expr=\thisrowno{0}*1e-6/#3, y index=1,col sep=space]
		{#1};
	\addlegendentry[mark options={scale=1}]{#2};
	}

\newenvironment{CompareSchedulersHist}[7][]
{
	\begin{tikzpicture} \tikzsetnextfilename{#1}
		\footnotesize
		\pgfplotstableset{format=file,header=false,col sep=tab}
		\pgfplotsset{STYLE plot/.style={
			width=130, height=120,
xlabel={\tiny \Mbps{}}, 
			ylabel={\tiny $P\{X \le x\}$}, y label style={at={(-0.12,0.5)}},
			title={#2},
			xmin=0,xmax=#5,
			every tick label/.append style={font=\tiny},
}}
		\begin{axis}[STYLE plot,]
			\ADDHIST{m4-data/m4.igr.MDV#3}{Ingress}{#4};

			\ADDHIST{m4-data/m4.cap.MW#3} {MW}{#4};
			\ADDHIST{m4-data/m4.cap.MDU#3}{MDU}{#4};
			\ADDHIST{m4-data/m4.cap.MDV#3}{MDV}{#4};

			\legend{} \draw [purple] (#6,-1) -- (#6,2);
			\draw [red] (#7,-1) -- (#7,2);
		\end{axis} 
	\end{tikzpicture}
}
{}

\newcommand{\ADDRATE}[6]{
	\immediate\write18{/Users/jerous/Documents/tools/transform/transform range #5 #6 <#1#2 >/tmp/data-#2.txt}
	\addplot+
		[restrict x to domain=#5:#6, restrict y to domain=0:600, mark options={scale=1},mark indices={1,60,120,180,240,360}]
		table[x index=0, y expr=\thisrowno{1}*1e-6/#4, col sep=space]
		{/tmp/data-#2.txt};
\addlegendentry[mark options={scale=1}]{#3};
	}

\newcommand{\SCENARIOCHAR}{}
\newcommand{\XMAXA}{0} \newcommand{\XMAXB}{0} \newcommand{\XMAXC}{0} \newcommand{\XMAXD}{0} \newcommand{\XMAXE}{0}
\newcommand{\RHOGA}{0} \newcommand{\RHOGB}{0} \newcommand{\RHOGC}{0} \newcommand{\RHOGD}{0} \newcommand{\RHOGE}{0}\newcommand{\RHOMA}{0} \newcommand{\RHOMB}{0} \newcommand{\RHOMC}{0} \newcommand{\RHOMD}{0} \newcommand{\RHOME}{0}

\newenvironment{CompareScenariosHistForGroupSec}[7][]
{
	\begin{figure}[ht!]
		\begin{subfigure}[b]{\SUBFIGWIDTH}
			\begin{CompareSchedulersHist}[scen#2B_#1]{#3}
				{Scenario5_paper_\SCENARIOCHAR2-0,repetition=0,user=0,group-sec=#1,range=0-100000,delay-pct=99.hist.txt}
				{#1}{\XMAXA}{\RHOGA}{\RHOMA}
			\end{CompareSchedulersHist}
		\end{subfigure}
		\begin{subfigure}[b]{\SUBFIGWIDTH}
			\begin{CompareSchedulersHist}[scen#2C_#1]{#4}
				{Scenario5_paper_\SCENARIOCHAR2-0,repetition=0,user=1,group-sec=#1,range=0-100000,delay-pct=99.hist.txt}
				{#1}{\XMAXB}{\RHOGB}{\RHOMB}
			\end{CompareSchedulersHist}
		\end{subfigure}
		\begin{subfigure}[b]{\SUBFIGWIDTH}
			\begin{CompareSchedulersHist}[scen#2D_#1]{#5}
				{Scenario5_paper_\SCENARIOCHAR2-0,repetition=0,user=2,group-sec=#1,range=0-100000,delay-pct=99.hist.txt}
				{#1}{\XMAXC}{\RHOGC}{\RHOMC}
			\end{CompareSchedulersHist}
		\end{subfigure}
		\begin{subfigure}[b]{\SUBFIGWIDTH}
			\begin{CompareSchedulersHist}[scen#2E_#1]{#6}
				{Scenario5_paper_\SCENARIOCHAR2-0,repetition=0,user=3,group-sec=#1,range=0-100000,delay-pct=99.hist.txt}
				{#1}{\XMAXD}{\RHOGD}{\RHOMD}
			\end{CompareSchedulersHist}
		\end{subfigure}
		\begin{subfigure}[b]{\SUBFIGWIDTH}
			\begin{CompareSchedulersHist}[scen#2_5_#1]{#7}
				{Scenario5_paper_\SCENARIOCHAR2-0,repetition=0,user=4,group-sec=#1,range=0-100000,delay-pct=99.hist.txt}
				{#1}{\XMAXE}{\RHOGE}{\RHOME}
			\end{CompareSchedulersHist}
		\end{subfigure}
		\caption{Scenario #2 ($w=#1 s$)}
		\label{fig:scen#2_#1}
	\end{figure}
} {}

\ifdefined\arxiv
	\newcommand{\affiliation}[1]{\begin{center}#1\end{center}}
	\newcommand{\email}[1]{#1}
	\newenvironment{keywords} {Keywords: \itshape} {}
	
\else
	
\fi

\usepackage{enumitem}
\usepackage{booktabs}
\usepackage{colortbl}

\newcommand{\TbrmFactor}{\exp ( \frac{\TBGTokens}{\TBGSigma} + \frac{\TBMTokens}{\TBMSigma} ) }
\renewcommand{\NI}{^n} 

\begin{document}

\title{Token Bucket-based Throughput Constraining in Cross-layer Schedulers}

\author{Jeremy Van den Eynde and Chris Blondia}

\ifdefined\arxiv
	\maketitle
	\affiliation{
		University of Antwerp - imec \\
		IDLab - Department of Mathematics and Computer Science \\
		Sint-Pietersvliet 7, 2000 Antwerp, Belgium \\
		\email{\{jeremy.vandeneynde,chris.blondia\}@uantwerpen.be}
	}
\else
	\affiliation{
		University of Antwerp - imec \\
		IDLab - Department of Mathematics and Computer Science \\
		Sint-Pietersvliet 7, 2000 Antwerp, Belgium \\
		\email{\{jeremy.vandeneynde,chris.blondia\}@uantwerpen.be}
	}
	\maketitle
\fi

\begin{abstract}
In this paper we consider upper and lower constraining users' service rates in a slotted, cross-layer scheduler context.
Such schedulers often cannot guarantee these bounds, despite the usefulness in adhering to \ac{QoS} requirements, aiding the admission control system or providing different levels of service to users.

We approach this problem with a low-complexity algorithm that is easily integrated in any utility function-based cross-layer scheduler.
The algorithm modifies the weights of the associated \acl{NUM} problem, rather than for example applying a token bucket to the scheduler's output or adding constraints in the physical layer.

We study the efficacy of the algorithm through simulations with various schedulers from literature and mixes of traffic.
The metrics we consider show that we can bound the average service rate within about five slots, for most schedulers.
Schedulers whose weight is very volatile are more difficult to constrain.
\end{abstract}

\begin{keywords}Cross-layer Scheduling, Quality of Service, Token Buckets, Resource allocation \end{keywords}

\footnotetext{Part of this research work was carried out at UAntwerpen, in the frame of Research Project FWO nr. G.0912.13 'Cross-layer optimization with real-time adaptive dynamic spectrum management for fourth generation broadband access networks'.
The scientific responsibility is assumed by its authors.}

\section{Introduction}
	In shared communication networks users often have to compete for service.
	For example, in wireless networks, the total available capacity is constantly fluctuating due to interference, non-stationary objects etc.
	Hence, users have to cooperate with each other to ensure the available capacity is shared fairly.

	Also in some wired network scenarios, such as \acs{DSL}, there is competition between users due to cross-talk \cite{verdyck2018network}.
	This cross-talk occurs in the copper cables when a user's signal leaks into other users' cables, thereby reducing the rate region, the set of all users' possible simultaneous data rates.
	This rate region is considered convex, implying that increasing one user's service rate, will decrease other users' rates.

	In these two settings, the physical layer has many Pareto-optimal operating points, which are all considered equally good to the physical layer.
	To the upper layers they are, however, not all equally good: allocations that satisfy the user's \ac{QoS} are preferable.
	For example, a live video feed has different requirements than video on demand, with regard to delay and data rates.
	Through cross-layering, the passing of information over the boundaries of the traditional OSI model layers, the upper layers can steer the physical layer.
	This cross-layering is often abstracted into a \ac{NUM} problem, described by \autoref{eq:NUM_problem}.

	\begin{equation} \label{eq:NUM_problem}
		\mathbf{\rate^*}=\argmax_{\mathbf{\rate} \in \RateRegion} \sum_{n=1}^N \Util(\Rate, \sysState)
	\end{equation}

	Here $n$ is the user index, $N$ the number of users, and utility function $\Util(\cdot)$ is an interface between the layers, representing the value of a user receiving a service rate $\Rate$.
	The state of the system $\sysState$ can include any observable information, such as queue lengths, arrivals, delays and others.
	We will omit $\sysState$ for readability when it is clear from the context.
	In general $\Util(\cdot)$ is an increasing, concave and differentiable function.
	Examples of cross-layer schedulers are mentioned in \autoref{sec:cxl}

	\autoref{eq:NUM_problem} chooses the operating point $\mathbf{\rate^*}$ from the rate region (all possible combinations of service rates) such that average utility over all users is maximized.
	In this system, a user $n$'s service rate depends on the weight in relation to all other users.
	This implies that it is challenging to control a single user's service rate, since there is no correspondence between one user's  isolated weight and its received rate.
	In spite of that, sometimes we want to be able to constrain the short-term (and long-term) average rates, for example to ensure the \ac{QoS} of the users.
	We give more reasons for implementing rate constraints in \autoref{sec:motivation}
	 
	Most cross-layer schedulers, such as the ones listed in \autoref{sec:cxl} do not offer the option to confine the service rate.
	Therefore, after reviewing the system model in \autoref{sec:model} we describe our contribution, the \ac{TBRM} algorithm, in \autoref{sec:algorithm}
	It is a generic low-complexity algorithm that constrains the short- and long-term average service rates of all users in a utility function-based cross-layer scheduler.
	Each time slot $t$, the \ac{NUM} problem is solved, but each user's weight is replaced by $\WeightB \TbrmFactor$.
		Two counters, $\TBGTokens$ and $\TBMTokens$, track the deficiency and excess in service, respectively.
		If a user $n$ has received less service than $\RhoG$, the amount of tokens will accumulate, and the user's weight will increase, therefore raising the probability of receiving more data rate.
		The parameters $\TBGSigma$ and $\TBMSigma$ are a measure for the slowness to react to deficiency and excess respectively.
		The variable $\WeightB$ accounts for non-positive weights.
	The algorithm is very easy to incorporate into any scheduler, as it does not require manipulating the original scheduler's weight function.

	In \autoref{sec:simulations} we evaluate our algorithm through multiple simulations.
	We compare our results with the unbounded scenarios, and look at the influence of the slot size $\tau$ and parameter $\sigma$.
	The metrics indicate that we can bound the average service rate for all users within a limited amount of time.
	Schedulers whose weight fluctuate heavily are more difficult to constrain.

	We close the paper with related works in \autoref{sec:related} and a conclusion in \autoref{sec:conclusion}.

\subsection{Motivation for service rate constraints} \label{sec:motivation}
	In a multi-user environment, it is useful to ensure that flows are guaranteed a bound on the average service rate.
	A provider might offer different \ac{QoS} guarantees to different users, depending on the subscribed model.
	This might include a guaranteed and/or maximal rate.

	There are other reasons to ensure a minimal rate.
	For example applications like audio and video need a minimal rate for a satisfying \ac{QoE}.
	Additionally, the authors of \cite{chakravorty2003flow} observe that TCP-based applications can lead to large queues when the throughput is too small.
	Finally, a guaranteed rate ensures that misbehaving competing flows, cannot smother flows from receiving their fair share.

	Applying an upper bound on a user's capacity is also useful.
	For example, to accommodate a new flow into a network, admission control algorithms often require an upper bound on the data rates \cite{qiu2001measurement}.
	If a flow disrespects this rate, other applications in the network can suffer deteriorated \ac{QoS}.
	By limiting the maximal data rate, a misbehaving flow is isolated and cannot negatively impact the other applications, but will only punish itself.
	In addition, it can also be useful to provide different service levels to users, where an operator may choose to cap the data rate for cheap data services, and remove this limit for the more expensive premium services.

	Although rate constraints can be implemented into the physical layer, it might be interesting to handle it at higher layers.
	First, it reduces the degrees of cross-layer freedom, and limits the communication necessary.
	Second, this allows a more flexible approach, allowing for temporary violations.
	Finally, it might not always be possible to introduce the rate constraints into the physical layer. 
	For example, the scheduler is implemented in hardware or closed-source and cannot be modified, or the corresponding \ac{NUM} problem's complexity might increase too much due to the additional constraints.

	Imposing upper bound constraints can be easily accomplished using a token bucket counter on a user's output stream.
	This is wasteful though: as the medium is shared, increasing one user's service rate means decreasing other users rates.
	By applying a token bucket, the excess capacity is thrown away as it is reserved by a particular user.

\subsection{Cross-layer schedulers} \label{sec:cxl}
	There are many existing cross-layer schedulers, each employing different metrics.
	We list here schedulers that are used in the simulations in \autoref{sec:simulations}

	Most schedulers found in literature are linear, meaning that the utility is of the form $u(\Rate)=\Weight \Rate$.
	For example, the \ac{MW} scheduler, presented in the seminal work \cite{tassiulas1992stability}, has $\Weight=\Queue$.

	The \acf{M-LWDF} \cite{andrews2001providing} uses $\Weight=\alpha\NI \HOL \frac{1}{\AvgRate}$, where $\alpha\NI=\frac{-\log(\qosPVio)}{\QOST}$, $\qosPVio$ is maximal delay violation probability, $\qosT$ the delay upper bound, $\HOL$ is the \acl{HOL} delay and $\AvgRate$ the exponentially averaged assigned data rates.

	The \ac{EXP/PF} \cite{basukala2009performance} scheduler, is a combination of Proportionally Fair scheduler and an exponent.
	It calculates the weights for real-time flows as $\exp(\frac{\alpha\NI \HOL - \chi}{1+\sqrt{\chi}}) \frac{1}{\AvgRate}$, where $\chi=\frac{1}{N} \sum_{n=1}^{N} \alpha\N \hol\N$.

	The \ac{MDU} scheduler \cite{song2005cross} employs the average waiting time:
	$\Weight=\frac{|u'^n(\AVG{U}\NI)|}{\AVG{\lambda}\NI}$, 
	where $u'^n$ is the derivative of the traffic class' utility function, $\AVG{U}\NI$ the average waiting time and $\AVG{\lambda}\NI$ the average arrival rate.

	The MD scheduler is the non-linear equivalent of the \ac{MW} scheduler, as it uses the utility function $u(\Rate)={\Queue}/{\Rate}$.\
	Finally, the \ac{MDV} scheduler \cite{van2017delay} is a non-linear scheduler of the form $u(\Rate)=-\frac{\Weight}{\Rate}$, where the weight $\Weight$ tries to minimize the amount of delay violations by looking at the queue and past delays.

\section{System model} \label{sec:model}
	Time in our model is divided into slots of $\tau$ seconds.
	There are $N$ users, indexed by $n \in \rangeII{1}{N}$, each of which can send $\tau \cdot \Capac(t)$ bits during slot $t \in \NatNum$, where $0 \le \Capac(t) \le \MaxCapac$ is the capacity for user $n$.

The user's arrivals and departures during slot $t$ are denoted by $A^n(t)$ and $D^n(t)$ respectively while the queue at the start of slot $t$ is indicated by $Q^n(t)$.
	The capacities $\VxCap(t)$ are determined by a scheduler, based on weights $\VxWeightT$:
	at the start of slot $t$, a request is made to the scheduler, the reply of which is applied at the start of slot $t+1$.
	There is thus a delay of $\tau$ seconds between a request and application of the rates.
	The scheduler takes the best matching rate from the rate region $\RateRegion \subset \mathbb{R}^N_+$.

	Each of the $N$ users has one traffic stream with delay upper bound $\UB{T}^n$, with the additional constraint that flow $n$'s average service rate is bounded: $0 \le \RhoG \le \Capac \le \RhoM \le \MaxCapac$.

\section{The Token Bucket Rate Modifier algorithm}\label{sec:algorithm}
\subsection{Token buckets}
Token buckets are found in various situations, such as 
	describing traffic flows \cite{tang1999network,cruz1991calculus,le2001network}, 
	to check conformance of incoming or outgoing traffic (policing and shaping \cite{stiliadis1998latency}), 
	traffic marking in DiffServ \cite{heinanen1999rfc,heinanen1999ietf} and 
	rate estimation \cite{zhang2015capacity}.

	Conceptually, a token bucket $TB(\rho,\sigma)$ consists of a bucket holding $\TBTokens$ tokens (e.g. bits).
	Tokens are added at a constant rate $\rho$ to the bucket, which is capped at $\sigma$ tokens.
	Whenever a packet of $B$ bit passes and there are sufficient tokens, $B$ tokens are removed from the bucket and the packet continues its journey.
	If $\TBTokens<B$, the packet is considered non-conforming and an appropriate action is taken, such as being color-marked non-conforming, queued (shaping) or dropped (policing).
	Such a token bucket will limit the long-term average outgoing rate to $\rho$.
	On a short-term scale, bursts of up to $\sigma$ bits can be served.

\subsection{Algorithm}
In the following algorithm, this token bucket principle is used to lower and upper bound the service rate in a cross-layer scheduler setting.
But, in contrast to a regular token bucket, we now do not cap the tokens to $\sigma$.
Rather, they are used to indicate the severity of the excess.

In the algorithm, instead of solving
\begin{equation}
	\argmax_{\mathbf{\rate} \in \RateRegion} \sum_n f(\Rate) \Weight
\end{equation}
where $f(\rate)=\rate$ for the linear, and $f(\rate)=\rate^{-1}$ for the reciprocal variant, the \ac{NUM} problem is modified to
\begin{equation} \label{eq:mod_NUM_problem}
	\argmax_{\mathbf{\rate} \in \RateRegion} \sum_n f(\Rate) \WeightB \TbrmFactor 
\end{equation}

Here, $\TBGTokens \in \rangeIE{0}{\infty}$ and $\TBMTokens \in \rangeEI{\text{-}\infty}{0}$ are the tokens for the guaranteed and maximal token buckets, respectively.
Every slot, the tokens are updated according to the following rules:

\begin{equation} \label{eq:tkg}
	\TBGTokens(t+1) = \max\{ 0, \TBGTokens(t)+ (\RhoG-\Capac(t)) \tau \}
\end{equation}
\begin{equation} \label{eq:tkm}
	\TBMTokens(t+1) = \min\{ 0, \TBMTokens(t)+ (\RhoM-\Capac(t)) \tau \}
\end{equation}

When the received capacity for a user $n$ in the past slots is less than the guaranteed rate $\RhoG$, the virtual token counter $\TBGTokens$ will continue to increase as long as there is a deficit in received service, and hence the weight will exponentially increase.
Likewise, if a user $n$ has received more than $\RhoM$ service, the virtual token counter $\TBMTokens$ will have a negative drift, as long as more data rate is assigned to the user.
This will reduce the user's weight exponentially.
When the service rate is less than $\RhoM$, the token counter will return to 0.

We introduce
\begin{equation}
	\WeightB = 
	\begin{cases}
		\AVG{\weight}, & \text{if}\ \Weight \le \epsilon\ \text{and}\ \frac{\TBGTokens}{\TBGSigma} + \frac{\TBMTokens}{\TBMSigma} \ne 0 \\
		\Weight, & \text{else}
	\end{cases}
\end{equation}
to account for non-positive weights $\Weight$.
Here $\AVG{\weight}$ is the exponentially averaged sum of all the positive weights, and $\epsilon$ a small number that will result in a capacity close to zero 
(for the simulations we used $\epsilon=\max_{n}\{\Weight\} \cdot 10^{-5}$).
If $\Weight \in \rangeEI{0}{\epsilon}$ it becomes difficult to increase the bandwidth reliably, and in the case of $\Weight = 0$, it is even impossible, since the weight will remain zero.
For a negative weight, which can occur for example for best effort flows in the EXP/PF scheduler, the additional factor would just result in an even lower weight, and would also inhibit us from receiving service.
$\AVG{\weight}$ is used to approximate a valid weight that is reasonably stable.
This weight is scheduler and traffic dependent, and thus must be calculated at run time.

Note that if $\RhoG=0$ or $\RhoM \ge \MaxCapac$, then respectively the first and second exponent will always be 1, and the respective bound is disabled.

It can be seen that if a flow stays within the bounds, then the tokens $\TBGTokens$ and $\TBMTokens$ will remain zero, and $\WeightB=\Weight$, resulting in the unmodified weight.
Only if some rate guarantee will not be met, weights will be adapted.

\subsection{Discussion} \label{sec:discussion}

\paragraph{Parameters $\RhoG$ and $\RhoM$}
The choice of $\rhoG$ and $\rhoM$ influences the speed at which the rate can adapt.
For example, if the guaranteed rate $\RhoG=0.75 \MaxCapac$, then in each slot the tokens can increase by at most $0.75\MaxCapac$, and the negative drift is at most $0.25\MaxCapac$.
Thus, if this flow has been receiving no service, then the tokens - and thus the weight too - will increase quickly.
If it is receiving service at a rate $\MaxCapac$, then the tokens will decrease more slowly.
A small $\RhoG$  thus also implies a small positive and large negative drift.
A similar reasoning can be applied to the maximal rate $\RhoM$.

\paragraph{Parameters $\TBGSigma$ and $\TBMSigma$}
In the traditional token bucket algorithms, $\sigma$ is a measure for the burstiness of a flow.
For example, large values of $\sigma$ mean that large bursts are allowed.
In our algorithm, the $\sigma$ can be interpreted as a measure for slowness to react.
A large value of $\TBMSigma$ means that longer periods of above-guaranteed service rates are possible, because our weight will decrease more slowly.
Small values of $\TBMSigma$ will react quicker and can lead to an overreaction.
The two token buckets can also influence each other: in case of an overreaction, the other token bucket will also have a sudden excess, and in turn have a fiercer reaction.
This can be observed for small values of $\sigma$ in the simulations of \autoref{sec:res-sigma}

\paragraph{Slot size}
In our system, the slot size implies a delay between a request for and subsequent assignment of the capacity.
A larger slot size means that changes will be slower, and that predicting future traffic becomes more important.
This also matters to the rate constraint algorithm, since the scheduler's response to weights becomes more unpredictable, hence modifying the weights.
The simulations of \autoref{sec:res-tau} briefly look at increasing slot sizes.

\paragraph{exp} The function $\exp$ is chosen here to modify the token fractions, but any continuous, strictly increasing function $\alpha(\cdot)$ for which holds that $\alpha(0)=1$, 
$\lim\limits_{x \rightarrow{-\infty}} \alpha(x)=0$ and 
$\lim\limits_{x\rightarrow\infty} \alpha(x)=\infty$ will give rate guarantees, albeit with different bounds.
Tests with different functions resulted in more short-time erratic behavior.

\paragraph{Additive form}
Instead of using a product, it is also possible to use an additive form, 
$$\argmax_{\mathbf{\rate} \in \RateRegion} \sum_n f(\Rate) \Weight
	+ (\alpha(\frac{\TBGTokens}{\TBGSigma})
	+ \alpha(\frac{\TBMTokens}{\TBMSigma})) \beta$$
Here $\alpha$ is a continuous, strictly increasing function with the properties 
	$\alpha(0)=0$, 
	$\lim\limits_{x \rightarrow{-\infty}} \alpha(x)=-\infty$ and 
	$\lim\limits_{x \rightarrow{ \infty}} \alpha(x)= \infty$.
An additional factor $\beta$ must be introduced to account for the fact that $\Weight$ is usually not unitless.

We ran some simulations for $\alpha(x)=x$ and $\alpha(x)=x^3$, and $\beta=\AVG{\weight}$. The simulations showed that this approach is also possible, and avoids the non-positive weight problem which forced us to introduce the factor $\WeightB$.
However, in the \ac{NUM} problem, what matters is the relative weights, rather than the absolute difference, which the additive form expresses.
Even though on larger timescales this leads to nicely averaged data rates, on short timescales the behavior is very extreme, where $\Capac(t)$ alternates between $0$ and rates close to $\MaxCapac$ in successive slots.

\paragraph{Complexity}
	The space and time complexity of the \ac{TBRM} algorithm is very low.
	Every slot we update the $N$ users' token counters $\TBGTokens$ and $\TBMTokens$ (Equations (\ref{eq:tkg}) and (\ref{eq:tkm})).
	Additionally, we have to select a suitable $\WeightB$ for all $n$.
	The exponentially weighted $\AVG{\weight}$ is a constant time operation $\BigOh(1)$.
	The resulting time complexity is thus $\BigOh(3N+1)=\BigOh(N)$. \newline
	Likewise, the space requirements are equally low: we track the $2 N$ counters, and a single exponentially weighted $\AVG{\weight}$.
	The space complexity is in this case $\BigOh(2N+1)=\BigOh(N)$.

\paragraph{Other considerations}
	Applying rate guarantees transforms a work-conserving scheduler into a non-work conserving scheduler.
	I.e. the scheduler might have capacity assigned, even though there are no jobs available.

	Additionally, throughput constraints reduces the stability region of a scheduler.
	Roughly, a scheduler is called stable for an arrival process if all the queues remain bounded.
	The set of arrival rates for which a scheduler is stable, is called the stability region.
	Schedulers such as \ac{MW} \cite{tassiulas1992stability} and \ac{MDU} \cite{song2005cross} have been proven to be stable for the widest range of arrivals.
	\newline
	Applying a constraint on the data rate for such schedulers reduces its stability region.
	Enforcing a maximal data rate inside the stability region, clearly decreases this region.
	Also supporting a minimal throughput constraint influences the stability region: a minimal throughput constraint can be rewritten as a (more complex) maximal throughput constraint on the other users.

\section{Simulations} \label{sec:simulations}

\subsection{Simulation setup}
We evaluated the \ac{TBRM} algorithm using simulations for the schedulers listed in \autoref{sec:cxl}
We ran simulations in OMNeT++ using the INET framework.
Every $\tau=50\text{ms}$ the original weights and weight modifiers were computed.
The resulting \ac{NUM} problem was then solved with the help of the nlopt \cite{nlopt} library, by first applying the local variant of the DIviding RECTangles algorithm \cite{gablonsky2001locally}, followed by the COBYLA algorithm\cite{powell1994direct}, to obtain the final rate, applying them in the next slot.

\iftrue
	\subsection{Rate region}
	The rate region was artificially generated by the following formula, which is based on the n-sphere formula.
	\begin{align*}
		r^n &=\MaxCapac \prod_{i=1}^{n-1} \sin(\phi^i)^{1-\gamma} \cos(\phi^n)^{1-\gamma}, r \in \rangeII{1}{N-1} \\
		r^N &=\capac^N_{max} \prod_{i=1}^{N-1} \sin(\phi^i)^{1-\gamma}
	\end{align*}

	For $N$ users, the rate region is the set of points generated by varying $\phi^i \in \rangeII{0}{\frac{\pi}{2}}, \forall i \in \rangeII{1}{N-1}$.
	The modifier $\gamma \in \rangeII{-1}{1}$ changes the shape of the rate region.
	If we set $\MaxCapac=M, \forall n$, then the shape ranges from an n-simplex for $\gamma=-1$, over an n-sphere with radius $M$ for $\gamma=0$, to a hypercube for $\gamma=1$.
\fi

\subsection{Scenarios}
The scenarios listed in \autoref{tab:scenarios} show the different types of traffic and the applied constraints.
\renewcommand{\arraystretch}{1.5}
	\colorlet{tableheadcolor}{gray!25} \colorlet{tablerowcolor}{gray!10} \newcommand{\rowcol}{\rowcolor{tablerowcolor}} \newcommand{\headcol}{\rowcolor{tableheadcolor}} \newcommand{\topline}{\arrayrulecolor{black}\specialrule{0.1em}{\abovetopsep}{0pt}\arrayrulecolor{tableheadcolor}\specialrule{\belowrulesep}{0pt}{0pt}\arrayrulecolor{black}}
	\newcommand{\midline}{\arrayrulecolor{tableheadcolor}\specialrule{\aboverulesep}{0pt}{0pt}\arrayrulecolor{black}\specialrule{\lightrulewidth}{0pt}{0pt}\arrayrulecolor{white}\specialrule{\belowrulesep}{0pt}{0pt}\arrayrulecolor{black}}
	\newcommand{\bottomline}{\arrayrulecolor{white}\specialrule{\aboverulesep}{0pt}{0pt}\arrayrulecolor{black}\specialrule{\heavyrulewidth}{0pt}{\belowbottomsep}}\newcommand{\bottomlinec}{\arrayrulecolor{tablerowcolor}\specialrule{\aboverulesep}{0pt}{0pt}\arrayrulecolor{black}\specialrule{\heavyrulewidth}{0pt}{\belowbottomsep}}

\begin{centering}
	\begin{table}[ht]
		\begin{tabular}{c  p{1.5cm}  p{1.5cm}  p{2cm}  p{1.5cm}  p{2cm} }
		\topline
		\headcol {Scenario} & {User 1} & {User 2} & {User 3} & {User 4} & {User 5} \\
		\midline

		1        & SAT \par[150,250] 
					& SAT \par[250,350] 
					& SAT \par[350,400]
					& SAT \par[150,350]
					& SAT \par[50,100]
					\\

		\rowcol
		2        & Starwars \par[50,150] 
					& Alice \par[250,350] 
					& \mbox{Self-Similar} \par[150,350]
					& SAT \par[150,350]
					& Sine2VS \par[50,120]
					\\
					
		3        & Starwars \par[50,150] 
					& Sine2F \par[250,350] 
					& Self-Similar  \par[150,350]
					& SAT \par[150,350]
					& Sine2VS \par[50,120]
					\\
					
		\rowcol
		4        & Sine2VS \par[150,250]
					& Sine2VS  \par[150,250]
					& Sine2VS \par[250,300]
					& Sine2VS  \par[150,350]
					& Sine2VS \par[50,400]
					 \\
					
		5        & Sine2VS  \par[150,250]
					& Sine2VS  \par[150,250]
					& Sine2VS  \par[250,300]
					& Sine2VS  \par[150,350]
					& Self-Similar \par[0,0]
					\\
		\bottomlinec
		\end{tabular}
		\caption[Summary of scenarios]{Summary of scenarios. Listed for each user are traffic type, and [$\rhoG$, $\rhoM$] in \Mbps{}.}
		\label{tab:scenarios}
	\end{table}
\end{centering}

The traffic types behave differently on short and large timescales.
The first type of traffic consists of a sine-wave, superimposed with a faster oscillating sine-wave. Some flows will oscillate slowly (Sine2VS) while others oscillate fast (Sine2F).
The second type of traffic is the heavy tail traffic, which is either a trace file of a video file, such as Starwars, or a self-similar flow, generated by a superposition of Pareto-distributed sources \cite{bai2013modeling}.
The last class of traffic, SAT, tries to send as much traffic as possible, by ensuring the queue is always backlogged.

These scenarios are run for $\tau=0.05s$ in \autoref{sec:res-regular}
In \autoref{sec:res-sigma} we vary $\TBSigma$, and in \autoref{sec:res-tau} it is $\tau$ that changes.

\subsection{Metrics}
We examined three different metrics.
The m2 and m3 metrics are defined on windows of size $G$, which groups $G$ consecutive slots.
\begin{enumerate}[label=m\arabic*]	
	\item the percentage of slots that would be marked non-conforming by a token bucket process 
		$TB(\rhoG, {\rhoG \tau x})$ and $TB(\rhoM, {\rhoM \tau x})$ for respectively the guaranteed and maximal rate.
		$x \in \mathbb{R}^+$ is a variable indicating the allowed burstiness.
		Increasing $x$ allows for more burstiness, and will result in a smaller percentage of non-conforming slots.

	\item 
		The average amount of excess bits per window $G$.
		If we define the amount of bit reserved in window $w$ as $C^G(w)=\sum_{t=wG}^{(w+1)G} C(t)\tau$, and $W$ as the total number of windows, then the m2 metric for respectively the guaranteed and maximal rate can be formally described as 
		$\langle \{ \max\{\rhoG G \tau-\capac^G(w), 0 \} | w=0 .. W-1 \} \rangle$ and
		$\langle \{ \max\{\capac^G(w)-\rhoM G \tau, 0 \} | w=0 .. W-1 \} \rangle$.
		This is a representation of the severity of the average violation. A larger number indicates more severe violations.
		The metric can be visualized by imagining the surface above or below the required rate.
		Increasing $G$ decreases the m2 metric as we smooth out excess bits over a larger window.

	\item $\langle B^G \rangle $: where $B^G$ is the set of consecutive violating windows.
		This metric gives an idea of how grouped violations are.
		For example, if this number is large, it means that a violation is resolved slowly.

\end{enumerate}

\subsection{Results}

\newcommand{\FIGWIDTH}{0.3\textwidth}

\newcommand{\PREFIX}{}
\newcommand{\XCOL}{} \newcommand{\XMIN}{} \newcommand{\XMAX}{} \newcommand{\XLABEL}{} \newcommand{\XMODE}{} 
\newcommand{\YCOL}{} \newcommand{\YMIN}{} \newcommand{\YMAX}{} \newcommand{\YLABEL}{} \newcommand{\YMODE}{} 
\newcommand{\LEGENDONLY} {}

\subsubsection{Regular scenarios} \label{sec:res-regular}
In the following plots, we averaged over all schedulers and scenarios, as showing the individual schedulers would result in a cluttered plot.
Important discrepancies between schedulers will be discussed in the text.

Each plot has two curves, one of which displays the results for which no rate constraints were applied, as a base case, and the other has our \ac{TBRM} algorithm applied.

\begin{figure}[ht] \renewcommand{\PREFIX}{regular.m5} 
		\renewcommand{\XLABEL}{Allowed burstiness} \renewcommand{\XMODE}{normal}

		\renewcommand{\YMIN}{0} \renewcommand{\YMODE}{normal}

		\centering
		
\tikzsetnextfilename{plotterlegend-avg}
\begin{tikzpicture}
	\footnotesize
    \begin{axis}[hide axis,
    xmin=10, xmax=50, ymin=0, ymax=0.4,
    legend style={draw=white!15!black,legend cell align=left,legend columns=-1}
    ]
	\definecolor{AVG}{HTML}{000000}
	\definecolor{MW}{HTML}{1B9E77}
	\definecolor{MLWDF}{HTML}{D95F02}
	\definecolor{EXPPF}{HTML}{E7298A}
	\definecolor{MDU}{HTML}{66A61E}
	\definecolor{MDV}{HTML}{7570B3}

\addlegendimage{line width = 1pt, mark=o,       color=MW	}   \addlegendentry{Normal};
	\addlegendimage{line width = 1pt, mark=+,       color=MLWDF	}  	\addlegendentry{With TBRM};
\end{axis} \end{tikzpicture} 
  \newline
		\begin{subfigure}[b]{\FIGWIDTH}
\ifdefined\arxiv
			\includegraphics{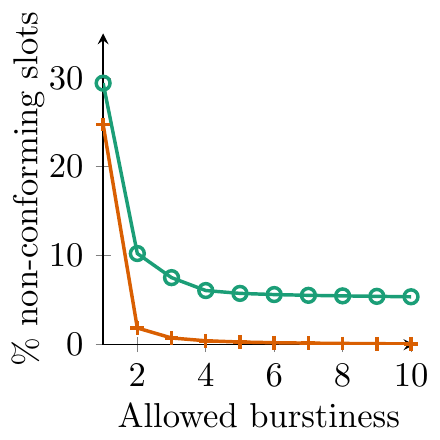}
\else
			\renewcommand{\XCOL}{\thisrow{sigma2mult}} \renewcommand{\XMIN}{1}  \renewcommand{\XMAX}{10.1}
			\renewcommand{\YCOL}{100*\thisrow{m5mmax}} \renewcommand{\YMAX}{35} \renewcommand{\YLABEL}{\MVYLABEL}
			\tikzsetnextfilename{regular-m5-max} \input{plotter-avg.tikz}
\fi
	        \vspace{-1.5\baselineskip}
			\caption{m1: upper bound on rate} \label{fig:regular-m5-max}
		\end{subfigure} \
		\hspace{10pt} \
		\begin{subfigure}[b]{\FIGWIDTH}
\ifdefined\arxiv
			\includegraphics{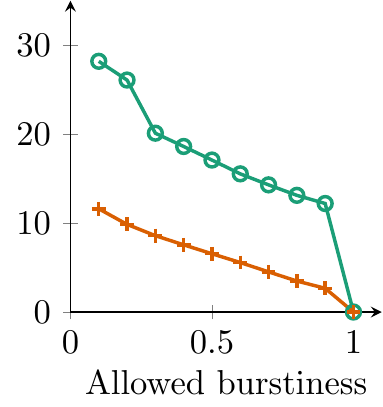}
\else
			\renewcommand{\XCOL}{0.1*\thisrow{sigma2mult}} \renewcommand{\XMIN}{0}  \renewcommand{\XMAX}{1.1}
			\renewcommand{\YCOL}{100*\thisrow{m5mmin}} \renewcommand{\YMAX}{35} \renewcommand{\YLABEL}{}
			\tikzsetnextfilename{regular-m5-min} \input{plotter-avg.tikz}
\fi
	        \vspace{-0.5\baselineskip}
			\caption{m1: lower bound on rate} \label{fig:regular-m5-min}
		\end{subfigure}
	    \vspace{-0.5\baselineskip}
		\caption{m1 for the regular scenarios} \label{fig:regular-m5}
	\end{figure}

\paragraph{m1}
	The m1 metric is shown in \autoref{fig:regular-m5}, which displays on the x-axis the allowed burstiness, and on the y-axis the percentage of non-conforming slots.

	If we examine \autoref{fig:regular-m5-max}, which shows the m1 metric for the upper bound, then we can see that for $x=1$, the number of violations is close to the results of the unconstrained simulations.
	Increasing $x$, the allowed burstiness, however, we can observe that the violation probability quickly drops for our \ac{TBRM} algorithm, and becomes almost 0 when the allowed burst size is $5 \rhoM \tau$.
	This indicates that the violations occur irregularly spread.
	The unconstrained results remain fixed around 6\% for a long time.

	The underlying data shows that for the constrained scenarios all the schedulers inhibit the same behavior: there is a steep decline in violations, going from $x=1$ to $x=2$, and then they gradually go to almost 0 for $x=5$.

	This behavior is the same for all schedulers over all traffic classes.
	However, the initial violation probability for SAT class is slightly lower than the video, self-similar and sine classes.
	The SAT traffic is easier to correct due to its queue based nature.

	In the m1 plot of the guaranteed rate in \autoref{fig:regular-m5-min}, our domain is limited to $\rangeEI{0}{1}$: if $x=1$, then it means that approximately in every slot we allow a deficit of $\rhoG \tau$ bit, which is obviously the maximum deficit we can attain per slot.
	In the plot, one can see that there is a much wider gap between the constrained and unconstrained scenarios, confirming the efficacy of our algorithm.

	The data shows here that the majority of the violations come from the \ac{MW} and \ac{M-LWDF} schedulers, and more specifically for the video streams.
	For example, in the \ac{TBRM} scenarios, for $x=0.1$, both schedulers have a violation probability of about 20\%, while the other schedulers are closer to 6\%.

	This difference can be explained by the fact that in those linear schedulers the queue length is used as a weight.
	This number is immediate, which causes a more unpredictable weight (especially in combination with a linear scheduler), making it more difficult to estimate a suitable weight modifier. 
	Additionally, the weight can become 0 very easily.
	This requires the use of the additional $\WeightB$ modifier. Even though $\AVG{\weight}$ is smoother, the switch between $\AVG{\weight}$ and $\Weight$ can also be disruptive.
	However, this extra factor is necessary, as simulations without this correction $\WeightB$, result in a much higher violation probability.

	The curve looks quite linear. This can be explained by the fact that the guaranteed rate violations are more evenly spread out.

\paragraph{m2}
	Ideally, we can limit the rate immediately.
	However, there is an inherent delay of 1 slot, and an elasticity in the form of a burst factor.
	Therefore, we study the rate, when we group $\frac{G}{\tau}$ slots into windows of size $G$.
	
\begin{figure}[ht] \renewcommand{\PREFIX}{regular} 
		\renewcommand{\XCOL}{\thisrow{groupsec}} \renewcommand{\XMIN}{0} \renewcommand{\XMAX}{2.1}
		\renewcommand{\XLABEL}{Window length (s)} \renewcommand{\XMODE}{normal}

		\renewcommand{\YMIN}{0} \renewcommand{\YMAX}{35} \renewcommand{\YMODE}{normal}

		\centering
		
\tikzsetnextfilename{plotterlegend-avg}
\begin{tikzpicture}
	\footnotesize
    \begin{axis}[hide axis,
    xmin=10, xmax=50, ymin=0, ymax=0.4,
    legend style={draw=white!15!black,legend cell align=left,legend columns=-1}
    ]
	\definecolor{AVG}{HTML}{000000}
	\definecolor{MW}{HTML}{1B9E77}
	\definecolor{MLWDF}{HTML}{D95F02}
	\definecolor{EXPPF}{HTML}{E7298A}
	\definecolor{MDU}{HTML}{66A61E}
	\definecolor{MDV}{HTML}{7570B3}

\addlegendimage{line width = 1pt, mark=o,       color=MW	}   \addlegendentry{Normal};
	\addlegendimage{line width = 1pt, mark=+,       color=MLWDF	}  	\addlegendentry{With TBRM};
\end{axis} \end{tikzpicture} 
  \newline
		\begin{subfigure}[b]{\FIGWIDTH}
\ifdefined\arxiv
			\includegraphics{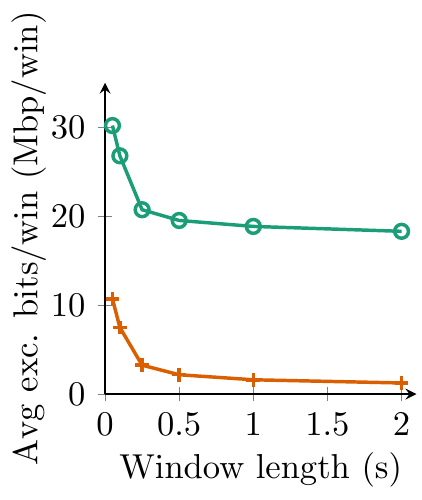}
\else
			\renewcommand{\YCOL}{\thisrow{m2mmax}} \renewcommand{\YLABEL}{\MIIYLABEL}
			\tikzsetnextfilename{regular-m2-max}\input{plotter-avg.tikz}
\fi
	        \vspace{-0.5\baselineskip}
			\caption{m2: upper bound on rate} \label{fig:regular-m2-max}
		\end{subfigure}\
		\hspace{10pt} \
		\begin{subfigure}[b]{\FIGWIDTH}
\ifdefined\arxiv
			\includegraphics{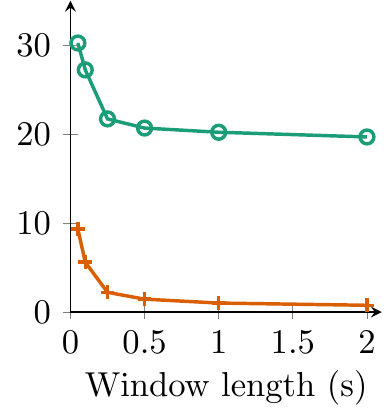}
\else
			\renewcommand{\YCOL}{\thisrow{m2mmin}} \renewcommand{\YLABEL}{}
			\tikzsetnextfilename{regular-m2-min}\input{plotter-avg.tikz}
\fi
	        \vspace{-0.5\baselineskip}
			\caption{m2: lower bound on rate} \label{fig:regular-m2-min}
		\end{subfigure}
	    \vspace{-0.5\baselineskip}
		\caption{m2 for the regular scenarios} \label{fig:regular-m2}
	\end{figure}

	The m2 metric in \autoref{fig:regular-m2}, displays the average amount of violated bits per window, for increasing window sizes $G$.

	It can be observed that for a window size of $G=\secnd{0.05}$, for the unconstrained scenarios there is an average of \Mbit{30}/window in excess of the target rate.
	When we constrain it using our algorithm, this drops to about \Mbit{10}/window.

	Increasing the window size $G$, averages out bursts.
	Like the results for m1, there is a steep decline until $G=\secnd{0.25}$, which coincides with 5 slots, after which the bit violations remains stable in the upper bound case.
	Though less pronounced, also here do the \ac{MW}, \ac{M-LWDF} and MD schedulers fare the worst for small window sizes, mainly for the Self-Similar traffic.

	The decline implies that bursts are usually short-lived: overflow and good windows are usually close together, as they don't violate the constraints when merged. This is also confirmed in the m3 metric, below.
	The rate of decline is similar for the scenarios with and without \ac{TBRM} applied.

\begin{figure}[ht] \renewcommand{\PREFIX}{regular} 
		\renewcommand{\XCOL}{\thisrow{groupsec}} \renewcommand{\XMIN}{0} \renewcommand{\XMAX}{2.1} 
		\renewcommand{\XLABEL}{Window length (s)} \renewcommand{\XMODE}{normal}

		\renewcommand{\YMIN}{0.5} \renewcommand{\YMODE}{log}

		\centering		
		
\tikzsetnextfilename{plotterlegend-avg}
\begin{tikzpicture}
	\footnotesize
    \begin{axis}[hide axis,
    xmin=10, xmax=50, ymin=0, ymax=0.4,
    legend style={draw=white!15!black,legend cell align=left,legend columns=-1}
    ]
	\definecolor{AVG}{HTML}{000000}
	\definecolor{MW}{HTML}{1B9E77}
	\definecolor{MLWDF}{HTML}{D95F02}
	\definecolor{EXPPF}{HTML}{E7298A}
	\definecolor{MDU}{HTML}{66A61E}
	\definecolor{MDV}{HTML}{7570B3}

\addlegendimage{line width = 1pt, mark=o,       color=MW	}   \addlegendentry{Normal};
	\addlegendimage{line width = 1pt, mark=+,       color=MLWDF	}  	\addlegendentry{With TBRM};
\end{axis} \end{tikzpicture} 
  \newline
		\begin{subfigure}[b]{\FIGWIDTH}
\ifdefined\arxiv
			\includegraphics{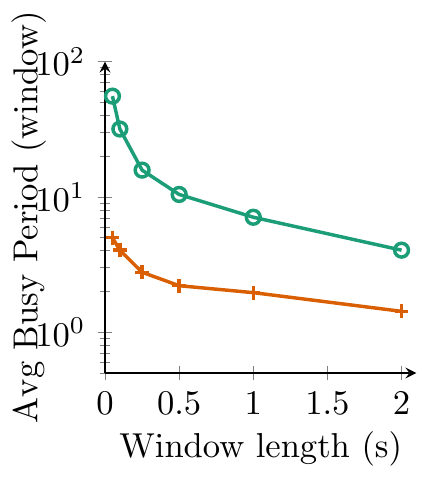}
\else
			\renewcommand{\YCOL}{\thisrow{m3mmax}} \renewcommand{\YMAX}{100} \renewcommand{\YLABEL}{\MIIIYLABEL}
			\tikzsetnextfilename{regular-m3-max}\input{plotter-avg.tikz}
\fi
	        \vspace{-0.5\baselineskip}
			\caption{m3: upper bound on rate} \label{fig:regular-m3-max}
		\end{subfigure}\
		\hspace{10pt} \
		\begin{subfigure}[b]{\FIGWIDTH}
\ifdefined\arxiv
			\includegraphics{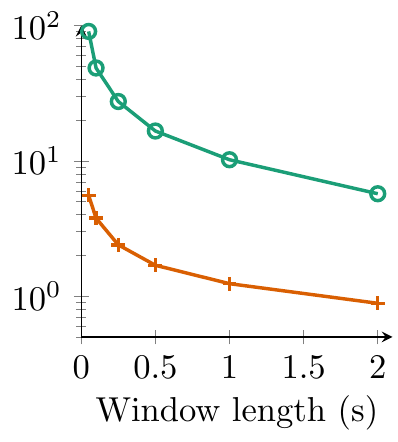}
\else
			\renewcommand{\YCOL}{\thisrow{m3mmin}} \renewcommand{\YMAX}{100} \renewcommand{\YLABEL}{}
			\tikzsetnextfilename{regular-m3-min}\input{plotter-avg.tikz}
\fi
	        \vspace{-0.5\baselineskip}
			\caption{m3: lower bound on rate} \label{fig:regular-m3-min}
		\end{subfigure}
	    \vspace{-0.5\baselineskip}
		\caption{m3 for the regular scenarios} \label{fig:regular-m3}
	\end{figure}

\paragraph{m3}
	The last metric discusses the average length of a violation streak.
	\autoref{fig:regular-m3} shows the average number of successive windows that violate their constraints in a log-plot.
	The m3 metric, like the m2 metric, initially decreases quickly as the window size increases, and then slowly decreases.
	It can be clearly seen that, regardless of the unconstrained behavior, the \ac{TBRM} algorithm limits the bursts to 5 windows, for $G=\tau$, dropping to 2 windows for $G=5\tau$.
	These short bursts confirm that the algorithm is able to fix excesses within about 5 slots.

	The main contributor to the average in this metric is the \ac{MDU} scheduler.
	The data show that this is because whereas other schedulers consist of many smaller busy periods, the \ac{MDU} scheduler has only one or two large busy periods, increasing the average significantly.

	Without the \ac{MDU} data, the average for 5 windows is about 1, for the minimal rate constraint, and 2 for the maximal rate constraint.

\subsubsection{Study of parameter \texorpdfstring{$\TBSigma$}{Sigma}} \label{sec:res-sigma}
\begin{figure}[ht]
		\renewcommand{\PREFIX}{sigma.m=5.m5} 
		\renewcommand{\XMODE}{log}
		\renewcommand{\YMIN}{0} \renewcommand{\YMODE}{normal}

		\centering
		
\tikzsetnextfilename{plotterlegend}
\begin{tikzpicture}
	\footnotesize
    \begin{axis}[hide axis,
	    xmin=10, xmax=50, ymin=0, ymax=0.4,
	    legend style={font=\scriptsize,draw=white!15!black,legend cell align=left,legend columns=-1}
    ]
	\definecolor{AVG}{HTML}{000000}
	\definecolor{MW}{HTML}{1B9E77}
	\definecolor{MLWDF}{HTML}{D95F02}
	\definecolor{EXPPF}{HTML}{E7298A}
	\definecolor{MDU}{HTML}{66A61E}
	\definecolor{MD}{HTML}{000000}
	\definecolor{MDV}{HTML}{7570B3}

\addlegendimage{line width = 1pt, mark=x,       color=MW   }   	\addlegendentry{MW};
	\addlegendimage{line width = 1pt, mark=+,       color=MLWDF}	\addlegendentry{MLWDF};
	\addlegendimage{line width = 1pt, mark=square,  color=EXPPF}	\addlegendentry{EXPPF};
	\addlegendimage{line width = 1pt, mark=triangle,color=MDU  }  	\addlegendentry{MDU};
	\addlegendimage{line width = 1pt, mark=*,		color=MD   }  	\addlegendentry{MD};
	\addlegendimage{line width = 1pt, mark=o,       color=MDV  }  	\addlegendentry{MDV};
\end{axis} \end{tikzpicture} 
 		\begin{subfigure}[b]{\FIGWIDTH}
\ifdefined\arxiv
			\includegraphics{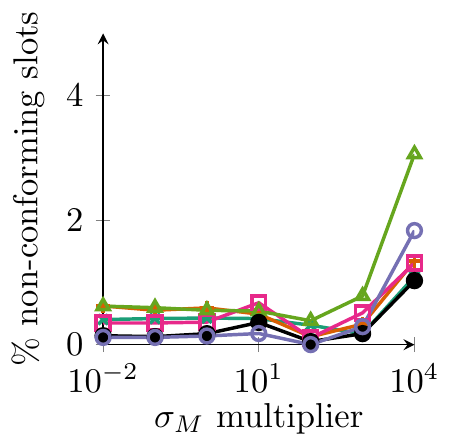}
\else
			\renewcommand{\XLABEL}{$\tbmsigma$ multiplier}
			\renewcommand{\XCOL}{\thisrow{sigmam}} \renewcommand{\XMIN}{1e-2}  \renewcommand{\XMAX}{1e4}
			\renewcommand{\YCOL}{100*\thisrow{m5mmax}} \renewcommand{\YMAX}{5} \renewcommand{\YLABEL}{\MVYLABEL}
			\tikzsetnextfilename{sigma-m=5-m5-max} \input{plotter.tikz}
\fi
	        \vspace{-1.5\baselineskip}
			\caption{m1: upper bound on rate} \label{fig:sigma-m=5-m5-max}
		\end{subfigure}\
		\hspace{10pt} \
		\begin{subfigure}[b]{\FIGWIDTH}
\ifdefined\arxiv
			\includegraphics{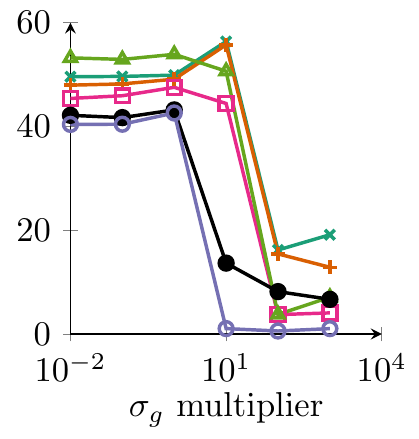}
\else
			\renewcommand{\XLABEL}{$\tbgsigma$ multiplier}
			\renewcommand{\XCOL}{0.1*\thisrow{sigmag}} \renewcommand{\XMIN}{1e-2}  \renewcommand{\XMAX}{1e4}
			\renewcommand{\YCOL}{100*\thisrow{m5mmin}} \renewcommand{\YMAX}{60} \renewcommand{\YLABEL}{}
			\tikzsetnextfilename{sigma-m=5-m5-min} \input{plotter.tikz}
\fi
	        \vspace{-0.5\baselineskip}
			\caption{m1: lower bound on rate} \label{fig:sigma-m=5-m5-min}
		\end{subfigure}
	    \vspace{-0.5\baselineskip}
		\caption{m1 for the $\sigma$ scenarios} \label{fig:sigma-m=5-m5}
	\end{figure}

	In this section we look at the results of varying burst parameter $\TBSigma$.
	We let $\tbgsigma=i\tau\rhoG$ and $\tbmsigma=i\tau\rhoM$, for $i \in \rangeII{10^{-2}}{10^4}$, and look at the effect on the m1 metric in \autoref{fig:sigma-m=5-m5}, which shows the results for the individual schedulers.

	The plot shows that for the upper bound constraints, the violation probability is always very low (about 4\% at most for $i=10^4$).
	The lower bound, however, starts around 50\% violation probability, and suddenly drops to smaller probabilities for $i=10^2$.

	Indeed, a small $\tbmsigma$ will overflow quickly, which causes its weight to be reduced swiftly, hence there will be less violations.
	As $\tbmsigma$ grows, the weight modifier will decrease much more slowly, leaving more room for violations.
	For the guaranteed rate, on the other hand, $\tbgtokens$ cannot build up a deficit as fast as $\tbmtokens$, as discussed before.
	It is only when the growths of the deficits are balanced that the m1 metric can lower, which is around $i=10^2$ and upwards.

	The schedulers that perform the worst are, unsurprisingly, the \ac{MW} and \ac{M-LWDF} schedulers.

\begin{figure}[ht]
	\renewcommand{\PREFIX}{tau.m=5.m5} 
	\renewcommand{\XLABEL}{$\tau$} \renewcommand{\XMODE}{normal}
	\renewcommand{\YMIN}{0} \renewcommand{\YMODE}{normal}

	\centering
	
\tikzsetnextfilename{plotterlegend}
\begin{tikzpicture}
	\footnotesize
    \begin{axis}[hide axis,
	    xmin=10, xmax=50, ymin=0, ymax=0.4,
	    legend style={font=\scriptsize,draw=white!15!black,legend cell align=left,legend columns=-1}
    ]
	\definecolor{AVG}{HTML}{000000}
	\definecolor{MW}{HTML}{1B9E77}
	\definecolor{MLWDF}{HTML}{D95F02}
	\definecolor{EXPPF}{HTML}{E7298A}
	\definecolor{MDU}{HTML}{66A61E}
	\definecolor{MD}{HTML}{000000}
	\definecolor{MDV}{HTML}{7570B3}

\addlegendimage{line width = 1pt, mark=x,       color=MW   }   	\addlegendentry{MW};
	\addlegendimage{line width = 1pt, mark=+,       color=MLWDF}	\addlegendentry{MLWDF};
	\addlegendimage{line width = 1pt, mark=square,  color=EXPPF}	\addlegendentry{EXPPF};
	\addlegendimage{line width = 1pt, mark=triangle,color=MDU  }  	\addlegendentry{MDU};
	\addlegendimage{line width = 1pt, mark=*,		color=MD   }  	\addlegendentry{MD};
	\addlegendimage{line width = 1pt, mark=o,       color=MDV  }  	\addlegendentry{MDV};
\end{axis} \end{tikzpicture} 
 	\begin{subfigure}[b]{\FIGWIDTH}
\ifdefined\arxiv
			\includegraphics{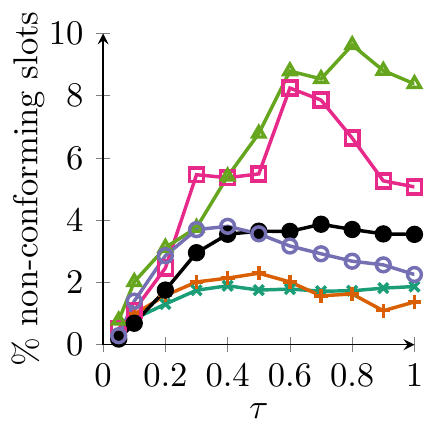}
\else
		\renewcommand{\XCOL}{\thisrow{tau}} \renewcommand{\XMIN}{0}  \renewcommand{\XMAX}{1}
		\renewcommand{\YCOL}{100*\thisrow{m5mmax}} \renewcommand{\YMAX}{10} \renewcommand{\YLABEL}{\MVYLABEL}
		\tikzsetnextfilename{tau-m=5-m5-max} \input{plotter.tikz}
\fi
        \vspace{-1.5\baselineskip}
		\caption{m1: upper bound on rate} \label{fig:tau-m=5-m5-max}
	\end{subfigure}\
	\hspace{10pt} \
	\begin{subfigure}[b]{\FIGWIDTH}
\ifdefined\arxiv
			\includegraphics{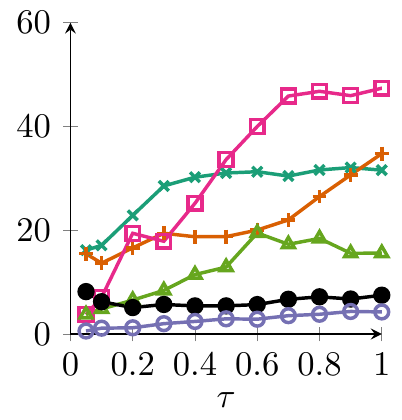}
\else
		\renewcommand{\XCOL}{\thisrow{tau}} \renewcommand{\XMIN}{0}  \renewcommand{\XMAX}{1}
		\renewcommand{\YCOL}{100*\thisrow{m5mmin}} \renewcommand{\YMAX}{60} \renewcommand{\YLABEL}{}
		\tikzsetnextfilename{tau-m=5-m5-min} \input{plotter.tikz}
\fi
        \vspace{-0.5\baselineskip}
		\caption{m1: lower bound on rate} \label{fig:tau-m=5-m5-min}
	\end{subfigure}
    \vspace{-0.5\baselineskip}
	\caption{m1 for the $\tau$ scenarios ($\sigma=5\tau\rho$)} \label{fig:tau-m=5-m5}
\end{figure}

\subsubsection{Study of parameter \texorpdfstring{$\tau$}{Slotlen}} \label{sec:res-tau}
	In the previous simulations, we assumed a slot length of $\tau=0.05s$.
	This study observes how the \ac{TBRM} algorithm changes in function of $\tau$.

	In \autoref{fig:tau-m=5-m5} the m1 metric for $\sigma=5\tau\rho$ is plotted.
	It can be observed that mainly the lower bound in  \autoref{fig:tau-m=5-m5-min} is sensitive to an increasing slot length.
	This might be because with an increasing $\tau$ also grows the probability of a larger delay: if in a slot a low capacity was assigned erroneously, the delays or queues will increase and additionally it takes longer to correct, causing larger queues.
	Especially the \ac{EXP/PF} scheduler suffers from this, as it is of the form $\exp(\hol)$, where $\hol$ is the \acl{HOL}.
	As the non-linear \ac{MD} and \ac{MDV} schedulers try to minimize the delay, they suffer less from an increase of slot size.

	For the upper bound in \autoref{fig:tau-m=5-m5-max} only the \ac{MDU} and \ac{EXP/PF} schedulers seem to suffer from the increased slot size.
	This is probably due to the fact that it is easier to receive a service lower than $\rhoM$.

\section{Related work} \label{sec:related}
	In \cite{andrews2001providing,shakkottai2001scheduling} the authors use virtual tokens as a measure for the average waiting time, and incorporate it with the \ac{M-LWDF}\cite{andrews2001providing} and \ac{EXP/PF}\cite{basukala2009performance} scheduling algorithm to warrant a minimal rate.
	It is, however, not transferable to other schedulers.
	In other schedulers, the guaranteed rate constraint is built into the scheduler itself \cite{mohseni2006optimized,wang2010stochastic}, but they are all scheduler-specific and don't allow enforcing a maximal data rate. 
	The authors of \cite{liu2003framework} consider utility based throughput allocation subject to certain properties, but is only valid for linear utility functions.
	In \cite{borst2003dynamic} a related problem of maintaining an optimal service rate is proposed.
	In \cite{andrews2005optimal}, the authors consider a generic algorithm with minimum and maximum rate constraints.
	It is, however, only applicable to schedulers that operate in function of an average rate.
	As such, it excludes for example the \ac{MW}\cite{tassiulas1992stability}, \ac{MD} and \ac{M-LWDF} schedulers.
	Other schedulers, such as, \ac{MDU}\cite{song2005cross} and \ac{MDV}\cite{van2017delay} have a more elaborate utility function and are more difficult to characterize.
	The authors of \cite{zahedi2018managing} also employ a token system, but assume that users lie about their demands to strategically maximize their utility.
	In \cite{Mandelli2019SatisfyingNS} constraints are applied to network slices of traffic aggregates in a 5G context, using an additive approach.

\section{Conclusion} \label{sec:conclusion}
	In this paper we looked at restricting the service rates given to users in a cross-layer scheduler setting.
	We implemented this using a low-complexity algorithm that modifies the weights in a \acl{NUM} problem, using the concept of token buckets.
	We first discussed cross-layer scheduling, and the need to both upper and lower limit data rates assigned to users.
	Then we proposed the \ac{TBRM} algorithm, and followed up with simulation results.
	We ran simulations for six different schedulers, and multiple scenarios demonstrating that using our approach it is possible to limit the service rate, within error, after about five slots for the maximal and guaranteed service rate for most schedulers.
	Schedulers that progress smoothly are easier to constrain than schedulers that can behave wildly, such as \ac{EXP/PF} for long slot times, \ac{MW} or \ac{M-LWDF}. 
	These are more difficult to restraint with respect to guaranteeing a lower bound on the service rate.

\ifdefined\arxiv
\bibliography{token_buckets.merged}
	\bibliographystyle{ieeetr}
\else
	\bibliography{shared-latex/all_refs.bib} 
	\bibliographystyle{shared-latex/IEEEtran}
\fi

\end{document}